\pdfoutput=1

\documentclass[11pt]{article}

\usepackage[final]{acl}

\usepackage{times}
\usepackage{latexsym}
\usepackage{booktabs}
\usepackage{adjustbox}
\usepackage{longtable}
\usepackage{array}
\usepackage{tcolorbox}
\usepackage{wrapfig}
\usepackage{graphicx}

\usepackage{authblk}

\title{Bahasa Harmony: A Comprehensive Dataset for Bahasa Text-to-Speech Synthesis with Discrete Codec Modeling of EnGen-TTS.}

\author{\textbf{Onkar Kishor Susladkar}}
\author{\textbf{Vishesh Tripathi}}
\author{\textbf{Biddwan Ahmed}}

\affil{\textbf{Yellow.ai}}

\affil{\texttt{\{onkar.susladkar,vishesh.tripathi,biddwan.ahmed\}@yellow.ai}}

\begin{document}
\maketitle
\begin{abstract}
This research introduces a comprehensive Bahasa text-to-speech (TTS) dataset and a novel TTS model, EnGen-TTS, designed to enhance the quality and versatility of synthetic speech in the Bahasa language. The dataset, spanning \textasciitilde55.0 hours and 52K audio recordings, integrates diverse textual sources, ensuring linguistic richness. A meticulous recording setup captures the nuances of Bahasa phonetics, employing professional equipment to ensure high-fidelity audio samples. Statistical analysis reveals the dataset's scale and diversity, laying the foundation for model training and evaluation. The proposed EnGen-TTS model performs better than established baselines, achieving a Mean Opinion Score (MOS) of 4.45 $\pm$ 0.13. Additionally, our investigation on real-time factor and model size highlights EnGen-TTS as a compelling choice, with efficient performance. This research marks a significant advancement in Bahasa TTS technology, with implications for diverse language applications. Link to Generated Samples: \url{https://bahasa-harmony-comp.vercel.app/}
\end{abstract}.

\section{Introduction}

The Bahasa language, spoken by a vibrant and diverse community, serves as a linguistic tapestry that encapsulates the rich cultural heritage of its speakers. In our increasingly digital world, the demand for advanced speech synthesis technologies tailored specifically to Bahasa becomes more pronounced. This necessity arises from the need for synthetic speech that authentically captures the nuances of Bahasa expressions, accommodating the linguistic diversity within the Bahasa-speaking population, including various dialects, registers, and cultural nuances.

Existing text-to-speech (TTS) systems, while making strides in the broader landscape, often fall short in addressing the requirements of Bahasa. This gap underscores the motivation for our research, which introduces a meticulously curated Bahasa TTS dataset and an innovative TTS model, EnGen-TTS. While other models have attempted to address the synthesis challenges for diverse languages, including Bahasa, they may exhibit drawbacks such as limited adaptability, linguistic richness, or efficiency.

Our proposed EnGen-TTS model not only fills these gaps but also showcases superior performance when compared to established baselines. The model achieves a remarkable Mean Opinion Score (MOS) of 4.45 $\pm$ 0.13, outperforming existing models even without fine-tuning on additional Bahasa data. Our key strength is, positioning EnGen-TTS as a solution for high-quality, adaptive Text-to-Speech synthesis across various languages.

\subsection{Contributions}

\begin{enumerate}
\item  Comprehensive Bahasa Dataset: Our research introduces a meticulously curated Bahasa text-to-speech (TTS) dataset, comprising \textasciitilde55.0 hours sourced from diverse linguistic contexts. This dataset, enriched with contributions from skilled voice artists and varied textual sources, stands as a valuable resource for the research community and addresses the need for a comprehensive linguistic foundation for Bahasa TTS systems. We will make dataset, trained-model and finetuneing code publicly available. Dataset Link: \url{https://bit.ly/3Vi22x9}

\item  Efficient Model Architecture: The proposed model architecture leverages a multi-lingual T5 (m-t5) encoder  \cite{xue-etal-2021-mt5} for confitioning text latents for decoding audio sequence through neural codec language modeling. This innovative design optimizes the synthesis process, allowing for finetuning more efficiently and reduced computational time while maintaining high-quality Bahasa speech synthesis.

\item Integration of Neural Codec Language Modeling: The incorporation of a trainable neural codec language modeling module represents a novel contribution. This module captures both textual and audio features, enhancing the model's ability to understand Bahasa linguistic nuances effectively. The integration of trainable weights in this module contributes to the adaptability and expressive power of the TTS system.

\item Versatile Pre-trained Model: Our research presents a pre-trained TTS model, EnGen-TTS, showcasing exceptional performance without additional fine-tuning on Bahasa-specific data. Trained on LJ-Speech  \cite{ljspeech17} and VCTK, this model exhibits inherent strengths in adapting to new languages, highlighting its versatility and potential for the development of high-quality Text-to-Speech systems for diverse languages like Bahasa.
\end{enumerate}

\section{Related Work}

In the domain of multilingual text-to-speech (TTS) datasets and models, several noteworthy contributions have been done for enhanced synthesis capabilities across diverse languages. The IndicSpeech: Text-to-Speech Corpus for Indian Languages \cite{srivastava-etal-2020-indicspeech} project recognizes the critical need for TTS systems tailored to the linguistic diversity of India. Presenting a 24-hour corpus for Hindi, Malayalam, and Bengali, the authors not only contribute data but also train state-of-the-art TTS systems for each language.

In a similar vein, the paper titled Towards Building Text-to-Speech Systems \cite{10096069} for the Next Billion Users explores the landscape of deep learning-based TTS systems, specifically focusing on the challenges and opportunities within the context of Indian languages. Some models like SLBERT \cite{susladkar2023slbert} a speech and language processing framework, which uses multimodal attention mechanism to get the better transition between the speech and language features. Recognizing the computational expense associated with investigating the multitude of Indian languages, lower resource availability, and untested advances in neural TTS, the authors evaluate various aspects such as acoustic models, vocoders, loss functions, training schedules, and speaker/language diversity. The results indicate that monolingual models with FastPitch  \cite{9413889} and HiFi-GAN V1  \cite{NEURIPS2020_c5d73680}, trained jointly on male and female speakers, exhibit significant improvements across 13 languages, as measured by mean opinion scores  \cite{VISWANATHAN200555}. 

Considering the landscape of existing TTS models, it's crucial to acknowledge advancements beyond the scope of the aforementioned papers. State-of-the-art models such as Tacotron  \cite{wang17n_interspeech}, WaveNet  \cite{45774}, and more recently, FastPitch and HiFi-GAN, have demonstrated significant progress in the realm of TTS. These models leverage deep learning architectures to generate natural and expressive speech, contributing to the evolving landscape of TTS technologies. 

\begin{table}[t]
\centering
\small
\begin{tabular}{@{}cc@{}}
\toprule
\textbf{Entity} & \textbf{Stats} \\ \midrule
\textbf{Hours} & \textasciitilde55 Hrs \\
\textbf{Mean Audio Length} & 4.06 Sec \\
\textbf{Total Words} & 458K \\
\textbf{Vocab Size} & 23K \\
\textbf{Sentences} & 68.9K \\
\textbf{Mean Word Freq.} & 9.4 \\
\textbf{Total Recordings} & 52K \\ \bottomrule
\end{tabular}
\caption{Descriptive statistics of our Bahasa corpus. We see that the corpus consists of a diverse vocabulary and is at a scale well-suited for state-of-the-art neural TTS models.}
\label{tab:my-table}
\end{table}

\section{Dataset}

In creating our Bahasa text-to-speech (TTS) dataset, we curated a linguistically diverse textual foundation. This dataset is integral to our innovative TTS model, EnGen-TTS. Drawing from sources like Wikipedia and incorporating content from chat-GPT translation, our approach involved a strategic gathering of text samples. This fusion of varied linguistic contexts lays the groundwork for a robust Bahasa TTS dataset, capturing the language's nuanced breadth of expression.

\subsection{Text collection}

The textual foundation for our Bahasa text-to-speech (TTS) dataset was meticulously curated from diverse sources, enriching the dataset with varied linguistic contexts. We gathered text samples from prominent repositories such as Wikipedia, ensuring a broad representation of topics and language styles. Additionally, we incorporated content generated through chat-GPT translation, further diversifying the dataset with conversational and translated expressions. This eclectic mix of sources contributes to a comprehensive and linguistically diverse textual corpus, laying the groundwork for a robust Bahasa TTS dataset.

\subsection{Speaker selection}

To imbue the TTS dataset with authentic and expressive voices, we engaged two skilled voice artists—one male and one female. These artists were selected based on their proficiency in Bahasa and their ability to convey the nuances of the language with clarity and naturalness. Both of the speakers are from southern Indonesia. The inclusion of both male and female voices ensures a balanced representation, catering to the diverse preferences of users interacting with the TTS system. The careful selection of voice artists contributes to the overall quality and authenticity of the recorded audio samples.

\subsection{Recording Setup}

Ensuring optimal recording quality is paramount for the success of any TTS dataset. Our recording setup was designed to capture the richness of Bahasa phonetics and nuances. A controlled acoustic environment was maintained to minimize external interference, and high-quality recording equipment was employed to capture the nuances of the voice artists' performances accurately. The setup included professional microphones, soundproofing measures, and studio-grade audio interfaces, creating an environment conducive to the production of high-fidelity Bahasa TTS audio samples. All the data we have recorded is at a sample rate of 48 kHz.

\subsection{Corpus Statistics}

We have a report of few statistics of our Bahasa Corpus in Table 1. Upon collecting the text data and organizing it into coherent sentences, the resultant Bahasa TTS corpus exhibits notable statistics reflecting the dataset's scale and diversity. The corpus comprises a total of 55 hours of recorded voice across 52K recordings. This extensive dataset is a testament to the effort invested in capturing a comprehensive range of linguistic expressions, ensuring the TTS system's adaptability to various applications and user preferences. These corpus statistics lay the foundation for subsequent model training and evaluation, fostering advancements in Bahasa TTS technology.

\subsection{Dataset Characterstics}

The dataset encompasses 52K recordings, featuring a vocabulary size of 23,000 unique terms. Comprising a total of 55 hours, evenly distributed between male and female speakers, we strategically allocated 5 hours from each speaker for validation and an additional 5 hours from each speaker for testing purposes.So a total of 10 hours of testing \& 10 hours of validation. This balanced selection ensures comprehensive coverage and representation in both the validation and test sets, fostering robust evaluation and training of our Bahasa TTS model, EnGen-TTS.

\begin{figure}[t]
    \centering
    \includegraphics[width=0.5\textwidth]{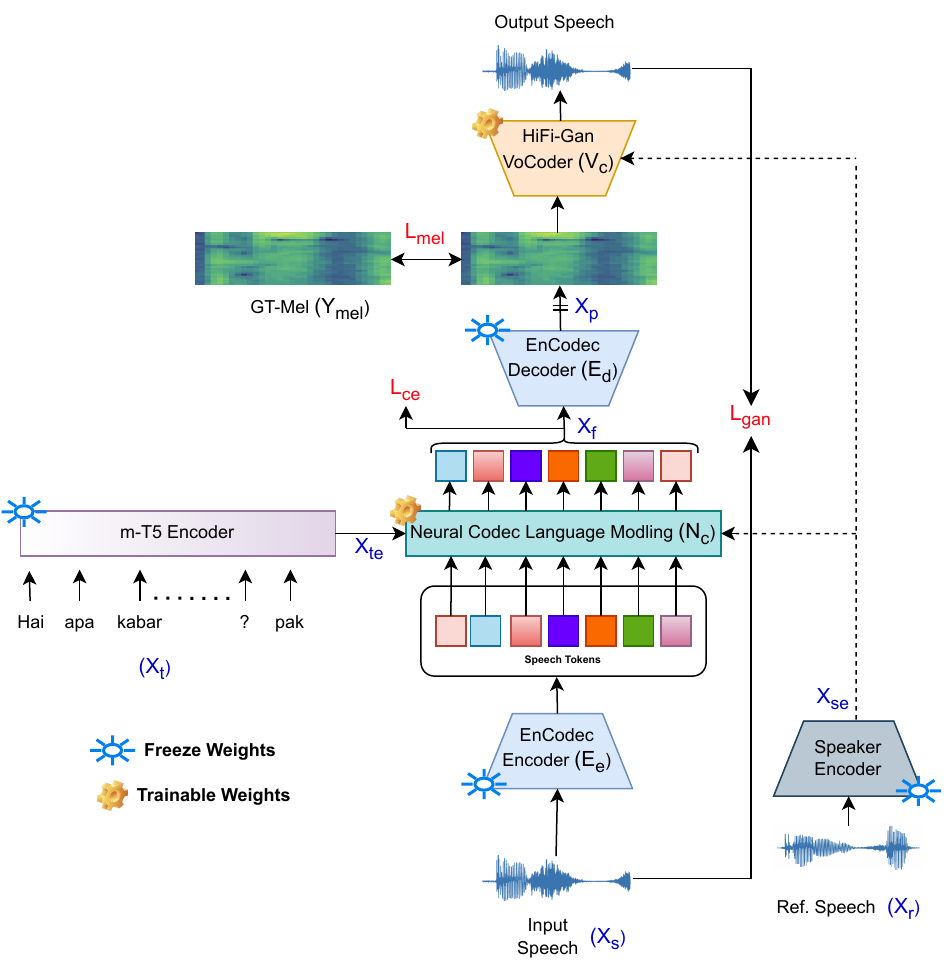}
    \caption{Architectural (EnGen-TTS) Framework for Bahasa Text-to-Speech Synthesis}
    \label{fig:architectue diagram}
\end{figure}

\section{Architecture}

Our research introduces the EnGen-TTS, a novel Bahasa text-to-speech synthesis system inspired by the state-of-the-art Encodec-based TTS Bark  \cite{schumacher2023enhancing}.The system leverages the architectural framework illustrated in Figure \ref{fig:architectue diagram}:

1. m-T5 Encoder: A frozen multi-li T5 encoder is utilized for conditioning on text latents. This encoder is pre-trained and is kept frozen during training. 

2. Audio Encodec: Audio Encodec from meta  \cite{kumar2024high} is pre-trained on an extensive 60K-hour audio dataset and kept frozen during training. This discretizes the audio into tokens, providing a robust audio representation without further training. 

3. Neural Codec Language Model: This module generates the audio sequence in an autoregressive manner. It conditions on both the text embeddings from the m-T5 encoder and the speaker embeddings, yielding a sequence that closely follows the linguistic and speaker-specific nuances of the input.

4. Speaker Encoder: A frozen encoder  \cite{wan2018generalized} trained on the LibriSpeech dataset. This module produces speaker latent vectors that condition the TTS output on the speaker's unique voice characteristics. 

5. HiFi-Gan Vocoder: It is for converting the mel spectrogram into natural speech, this vocoder is fine-tuned to adapt to the specific frequency profiles of the Bahasa language. The system is designed to synthesize natural-sounding Bahasa speech by conditioning on both linguistic content and speaker identity.

\textbf{Audio Codec Setting:} We adopt the pre-trained EnCodec model\cite{kumar2024high} as our tokenizer, a convolutional encoder-decoder model handling 22050 Hz audio at variable bitrates. The encoder generates embeddings at 75 Hz for 22050 kHz input, reducing the sampling rate by 320 times. These embeddings use residual vector quantization (RVQ) with 4 hierarchical quantizers of 1024 entries each, corresponding to a 3K bitrate for audio reconstruction at 22050 Hz.bFor our purposes, we use only the first entity of the 750 x 4 discrete representation matrix, as it contains all phonetic and content information, resulting in a 750 x 1 matrix. Higher bitrates, such as 12K, require more quantizers (e.g., 16) and offer better reconstruction quality. The EnCodec decoder then reconstructs the waveform at 22050 Hz from the discrete codes.

\subsection{Method}

Let, $X_s$ be the input audio, $X_t$ is the text corresponding to $X_s$, and, $X_r$ be the reference audio of the same speaker. The methodology commences with byte pair encoding (BPE) to convert $X_t$ into input IDs. These are fed into the frozen m-T5 encoder to derive text embeddings $X_{te}$. Concurrently, $X_s$ is discritized by the frozen Audio-Encodec encoder $E_{e}$, producing discrete audio tokens in between (0 to 1023). The Speaker Encoder processes $X_r$  to generate a speaker latent vector for each frame. These vectors ($X_{se}$) are then used to condition the model on the speaker's voice, providing a direct sequence that correlates with the length of the $X_r$ sample. Then, Neural Codec Language Modeling $N_{c}$ predicts the audio sequence conditioned on $X_{te}$ and $X_{se}$. The Sequential cross-entropy loss ($L_{ce}$) between the generated audio sequence and the ground truth audio sequence is computed here to ensure the fidelity of the audio tokens $X_{f}$. The predicted audio sequence is then passed through the Encodec Decoder $E_{d}$ to produce an intermediate audio representation. We calculate the loss ($L_{mel}$) between the predicted mel-spectrogram $X_{p}$ and the GT-Mel ($y_{mel}$) to ensure the model accurately captures the handcrafted audio features.

$L_{ce} = -\log(N_{c}(E_{e}(X_s), X_{te}, X_{se})))$


$X_{f} = N_{c}(E_{e}(X_s), X_{te}, X_{se}))$


$L_{mel} =  - \mid y_{mel} - E_{d}(X_{f}) \mid$


At last, To achieve natural-sounding audio, we pass intermediate udio representation to the HiFi-Gan vocoder $V_{c}$ which is conditioned on the speaker embeddings $X_{se}$ to prevent mode collapse. The loss ($L_{gan}$) between the predicted speech and input speech is computed to align the output with the input audio distribution. To compute the Gan loss we use the temporal Discriminator module as a $D$. The loss follows:


  $L_{gan} = -\log(D(V_c(E_d(X_f), X_{se}))) + | X_s - V_c(E_d(X_f), X_{se}) |$

\begin{figure}[!t]
    \centering
    \includegraphics[width=0.45\textwidth]{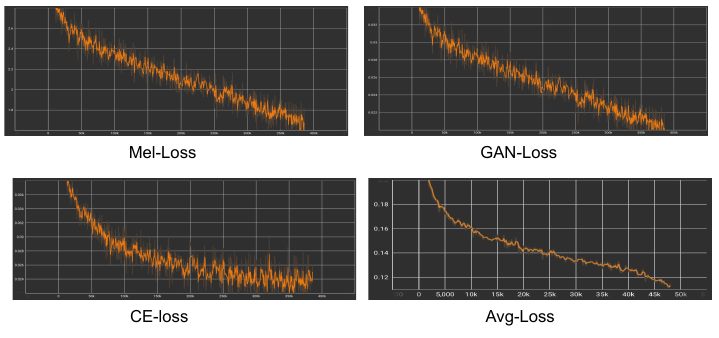}
    \caption{Loss plots}
    \label{fig:loss_plots}
\end{figure}

We adopt a composite loss function, taking a weighted average of $L_{ce}$, $L_{mel}$, and $L_{gan}$ to update the model's weights. This combination of losses ensures that the model learns not only the accurate prediction of audio tokens but also the refined generation of  melspectrograms and the final audio output. From Figure \ref{fig:loss_plots} we can see that all the 3 losses are helping model to learn optimally. The overall loss ($L_t$) is expressed as follows:

$L_t = \alpha L_{ce} + \beta L_{mel} + \gamma L_{gan}$

From our experiment, we determined the optimal values for the coefficients as follows: $\alpha=1.2$, $\beta=0.7$, $\gamma=0.6$.

\subsection{Audio Codec language modeling}

\begin{figure}[!t]
    \centering
    \includegraphics[width=0.5\textwidth]{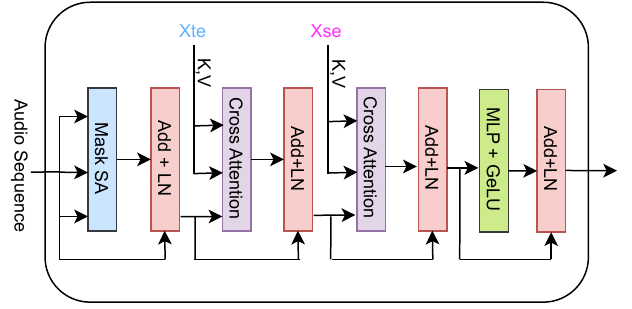}
    \caption{Audio Codec language modeling module}
    \label{fig:audio_sequence}
\end{figure}

In this section, we delve into the core component of our model, namely the Audio Codec Language Modeling Module. This pivotal module encompasses a Masked Self-Attention mechanism, Layer Normalization, two Cross-Attention blocks, and FeedForward layers, all integrated with GeLU non-linearity. As depicted in Figure \ref{fig:audio_sequence}, the process begins with discrete audio tokens being right-shifted and passed through the embedding layer to obtain audio embeddings. These embeddings are then directed through the Masked Self-Attention block. This specific block is utilized to decode the audio sequence autoregressively, akin to the GPT model, where the full scope of tokens isn't available during decoding. Utilizing normal self-attention could lead to overfitting on particular sequences and a lack of generalization across different audio sequences. The output from the masked self-attention undergoes layer normalization, augmented with a residual connection from the input audio embeddings. Subsequently, the embeddings are processed through a cross-attention module, conditioned on the text embeddings ($X_{te}$) derived from the m-T5 encoder. The output from this cross-attention phase is then subject to another layer normalization, followed by a residual connection from the preceding text-conditioned cross-attention block. 

In contrast to previous methods \cite{zhang2023speak} that concatenate speaker embeddings directly with audio embeddings, our approach employs an additional cross-attention step for conditioning on speaker embeddings ($X_{se}$), enhancing the model’s generalizability across multiple speakers in TTS applications. This output is finally channeled to a feedforward layer, followed by GeLU non-linearity and layer normalization.
The architecture maintains a consistent embedding dimension of 1024 and 16 attention heads across all sub-modules. During inference, we also implement a KV-cache mechanism to enhance efficiency. Our larger model configuration comprises 26 such blocks, each meticulously designed to optimize performance and accuracy in audio processing tasks.

\subsection{Inference}

During inference, the Neural Codec Language Model (Nc) is initialized with the <SOS> (start of sequence) token as the input. The model then autoregressively generates the entire sequence, conditioned on the text embeddings produced by the pre-trained m-T5 model and the speaker embeddings derived from the specified reference audio. During training, audio codec inputs help the model learn the mapping between audio and text embeddings. However, during inference, the model relies on learned embeddings and the autoregressive mechanism to generate the sequence. Our current training does not use scheduled sampling or a weaning-off period typical of teacher forcing. The model smoothly transitions from training to inference due to robust text and speaker conditioning.

\section{Experiments}

\subsection{Experimental Setup}
We assessed our Text-to-Speech (TTS) models quality using Mean Opinion Score (MOS) and Comparative Mean Opinion Score (CMOS). MOS measures average listener preferences, indicating the models' naturalness, pleasantness, and intelligibility. CMOS compares models directly, identifying slight differences in perceived quality. Together, these metrics offer detailed insights into each model's performance and real-world applicability.

\subsection{Real-Time Factor (RTF)}

It is a measure of how quickly a TTS system can generate speech relative to the length of the input text. Specifically, the RTF is calculated as the ratio of the time taken to synthesize speech to the duration of the resulting audio. For example, an RTF of 1.0 means that the TTS system takes one second to generate one second of speech. An RTF of less than 1.0 indicates that the system is faster than real-time, whereas an RTF greater than 1.0 indicates that the system is slower than real-time. Lower RTF values are generally preferred as they indicate a more efficient and faster TTS system.

\subsection{Quantitative Results}

\begin{table}[h]
\centering
\begin{adjustbox}{max width=\columnwidth} 
\begin{tabular}{@{}clrr@{}}
\toprule
\textbf{Baseline} & \multicolumn{1}{c}{\textbf{MOS}} & \multicolumn{1}{c}{\textbf{CMOS}} & \multicolumn{1}{c}{\textbf{RTF}} \\ \midrule
GradTTS \cite{popov2021grad} & 4.01 $\pm$ 0.11 & -0.0515 & 0.019 \\
GlowTTS \cite{kim2020glow} & 4.16 $\pm$ 0.14 & -0.0422 & 0.031\\
VITs \cite{kim2021conditional} & 4.19 $\pm$ 0.12 & -0.0366 & 0.023\\
NaturalSpeech \cite{tan2024naturalspeech} & 4.39 $\pm$ 0.19 & -0.0201 & 0.014 \\
XTTS \cite{casanova2024xtts} & 4.39 $\pm$ 0.12 & -0.0222 & 0.013 \\
CLaM-TTS \cite{kim2024clam} &	4.41 $\pm$ 0.09 & -0.0189 & 0.027 \\
VAENAR-TTS \cite{lu2021vaenar} & 4.36 $\pm$ 0.21 & 	-0.0326  & 0.014 \\
EnGen-TTS-L (Without pretrained) & 4.41 $\pm$ 0.10 & -0.0161 & 0.016\\
EnGen-TTS-L & \textbf{4.45 $\pm$ 0.13} & \textbf{-0.0101} & \textbf{0.016}\\ 
\bottomrule
\end{tabular}
\end{adjustbox}
\caption{Comparison of all models on Bahasa Datasets}
\label{table:results}
\end{table}

As the Bahasa language is written in Latin script, So there is always phonetic misalignment between speech and the input text. To learn those alignments we first trained our models and baselines on LJ-speech\cite{ljspeech17} and VCTK\cite{Yamagishi} dataset. After pertaining we use these learn weights for fine-tuning on our proposed bahasa dataset. We found that  Our EnGen-TTS-L (Large) model outperforms other previous baselines at  Bahasa text-to-speech synthesis. We Evaluate our model and baseline on MOS and CMOS metrics from Table \ref{table:results}  Our EnGen-TTS-L model achieves the highest MOS of 4.45 $\pm$ 0.13, surpassing all other models in the comparison. This indicates that listeners rated the speech generated by EnGen-TTS-L as more natural and closer to human speech than that of the competing models. Notably, EnGen-TTS outperforms NaturalSpeech and CLaM-TTS, which have MOS scores of 4.39 $\pm$ 0.19 and 4.41 $\pm$ 0.09, respectively. 

Even without pre-training on LJ-speech and VCTK dataset, our Engen-TTS-L model achieves results comparable to NaturalSpeech, CLaM-TTS, and XTTS, shows EnGen-TTS-L is good at adapting to new languages, which is important for creating high-quality speech synthesis for languages like Bahasa.

In terms of CMOS EnGen-TTS also exhibits superior performance. It achieves the lowest (best) score in Metric A with -0.0101, indicating a closer alignment with target speech characteristics than other models.

We also evaluate baselines and our model on the RTF (real-time factor) for generating the speech. From Table \ref{table:results} we found that EnGen-TTS is comparable fast, producing speech almost in real-time.  The EnGen-TTS is not only fast, but it also produces high-quality speech unlike Other methods, GradTTS and GlowTTS, are a bit slower and don't generate speech quality as well as EnGen-TTS. Our findings are important for people who want to produce speech systems and want the real-time inference with high and robust quality of speech.

These results collectively affirm that EnGen-TTS not only advances the state-of-the-art in TTS by delivering the most natural-sounding speech but also maintains high performance across various evaluation metrics. Which highlights the strength of our model's architecture and training methodology.

\subsection{Quntiative Results based on Multi-Lingual Dataset}


\begin{table*}
\centering
\begin{adjustbox}{max width=\textwidth}
\begin{tabular}{@{}ccccc@{}}
\toprule
\textbf{Languages} & \textbf{No of Hours} & \textbf{VITS} & \textbf{NaturalSpeech} & \textbf{EnGen-TTS-L} \\ \midrule
Spanish    & 23.34 & 3.39 & 3.72 & \textbf{4.28} \\
Portuguese & 17.82 & 3.41 & 3.78 & \textbf{4.37} \\
German     & 18.03 & 3.27 & 3.47 & \textbf{4.17} \\
Dutch      & 29.76 & 3.18 & 3.31 & \textbf{4.14} \\
Hindi      & 15.11 & 3.56 & 3.88 & \textbf{4.55} \\
Marathi    & 09.07  & 3.92 & 4.01 & \textbf{4.87} \\
Tamil      & 10.56 & 3.88 & 3.96 & \textbf{4.78} \\\bottomrule
\end{tabular}
\end{adjustbox}
\caption{Comparison of MOS scores for different languages using VITS, NaturalSpeech, and EnGen-TTS-L models.}
\label{table-langs}
\end{table*}

We have trained the model in 7 languages, For Latin languages (Spanish, Portugeas, German, and Dutch) we used the CML tts-dataset \cite{oliveira2023cml} and for indic languages (Hindi, Marathi, and Tamil), we used the indic-speech \cite{srivastava2020indicspeech} dataset. For Fair comparison, We loaded the models that are pre-trained from LjSpeech and VCTK datasets. We evaluate each model with the MOS score. We with our EnGen-TTS-L, we used two more baselines to showcase our novelty. 

The Table \ref{table-langs} presents a comparison of Mean Opinion Scores (MOS) across different languages for three models: VITS, NaturalSpeech, and EnGen-TTS-L. The languages evaluated include Spanish, Portuguese, German, Dutch, Hindi, Marathi, and Tamil, with varying amounts of training data for each language. EnGen-TTS-L consistently outperforms both VITS and NaturalSpeech across all the languages evaluated. For example, in Spanish, EnGen-TTS-L achieves a MOS of 4.28, significantly higher than both VITS (3.39) and NaturalSpeech (3.72). Similarly, in Portuguese, EnGen-TTS-L scores 4.37, surpassing VITS (3.41) and NaturalSpeech (3.78).

The performance gap is especially pronounced for languages like Marathi and Tamil, where EnGen-TTS-L achieves the highest scores of 4.87 and 4.78, respectively. In contrast, NaturalSpeech performs noticeably worse with MOS scores of 4.01 and 3.96 for Marathi and Tamil, respectively, while VITS scores lower at 3.92 for Marathi and 3.88 for Tamil.

Notably, EnGen-TTS-L also performs exceptionally well in Hindi with a MOS of 4.55, significantly outperforming both VITS (3.56) and NaturalSpeech (3.88). In German and Dutch, which have relatively more training data, EnGen-TTS-L continues to lead with MOS scores of 4.17 and 4.14, respectively, demonstrating consistent high-quality speech synthesis across languages, even for those with limited training data.

Overall, the results clearly indicate that EnGen-TTS-L excels in generating more natural-sounding speech across multiple languages, especially when compared to the baseline models, VITS and NaturalSpeech. This demonstrates the robustness of the EnGen-TTS-L model in multilingual TTS tasks, even with varying amounts of training data for each language.

\begin{figure}[!t]
    \centering
    \includegraphics[width=0.5\textwidth]{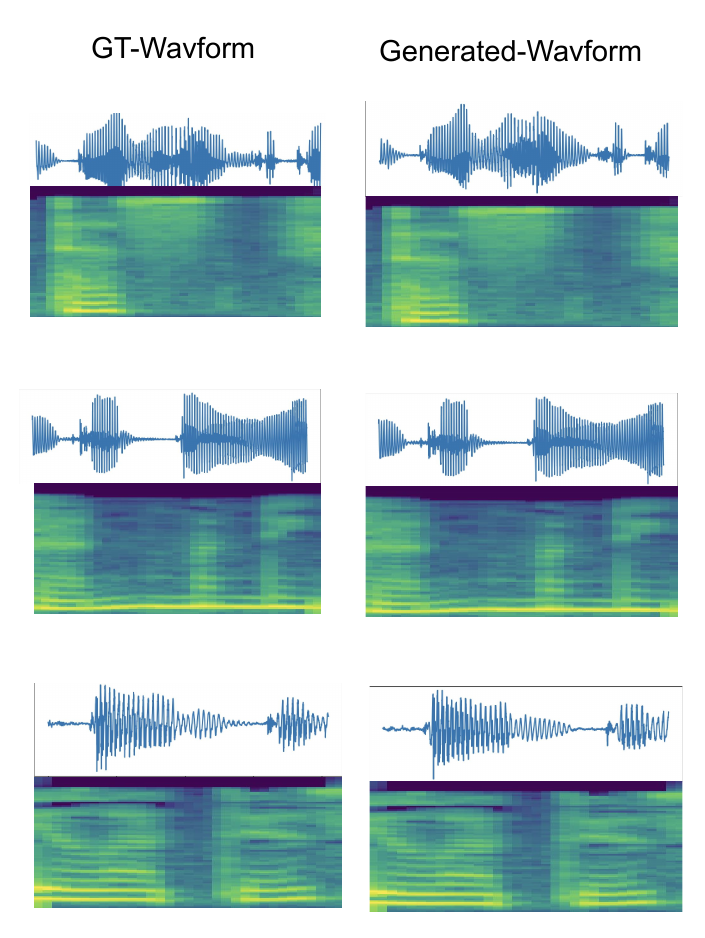}
    \caption{Comparison between Ground truth and Generated Audio}
    \label{fig:mel_wav}
\end{figure}

\subsubsection{Model Performance}

The model generates audio at a 24 kHz quality. When it comes to pronouncing acronyms, the task can be challenging. A helpful strategy is to articulate each letter separately, spacing them out to improve clarity. For numerical data, converting digits into their word equivalents often yields better results. An important observation is that the model might inadvertently replicate the reference speaker's audio in its output, particularly when the input text closely mirrors the reference material. The overall quality of the output is heavily influenced by the caliber of the reference audio. Ideally, the reference should be between 4 to 6 seconds long and exclusively contain clear speech, free from any background noises. It's worth noting that employing a cartoon-like voice in audio references might lead to model failure, as such inputs are significantly different from the data used during the training process. The model's capacity is constrained to 604 audio tokens and 1024 text tokens, where 600 audio tokens equate to approximately 16 seconds of sound. Look in Figure \ref{fig:mel_wav} for a comparison between the Ground truth waveform and the EngenTTS Generated waveform. We can see that our model-generated output is very close to the ground truth.

\subsubsection{Implementation details}

The EngenTTS-L's Audio Codec language modeling module utilizes transformer architecture with 26 blocks, 16 attention heads, a hidden dimension of 1024, a feed-forward layer dimension of 1024. The average length of the waveform in LJspeech and VCTK is 9.8 seconds, for our Bahasa dataset average length of the audio is around 7 seconds. During training, we randomly crop the waveform to a random length between 2 seconds and 6 seconds. Its corresponding phoneme alignments are used as the phoneme prompt. We remove the consecutive repetitions in the force-aligned phoneme sequence. While training we keep max sequence length of 500. The models are trained using 3 NVIDIA RTX 3090Ti 24 GPUs with a batch size of 4 with gradient accumulation steps of 24 per GPU for 800k steps. We optimize the models with the AdamW optimizer, warm up the learning rate for the first 32k updates to a peak of $5 \times 10^{-4}$, and then linear decay it.

\begin{table*}
\centering
\begin{adjustbox}{max width=\textwidth}
\begin{tabular}{@{}ccccccc@{}}
\toprule
\textbf{Model} & \textbf{Parameters} & \textbf{MOS} & \textbf{Hidden Dim} & \textbf{Attention Heads} & \textbf{No. of Blocks} & \textbf{RTF} \\ \midrule
EnGen-TTS-S & 87M & 4.12 $\pm$ 0.19 & 512 & 4 & 6 & 0.014 \\
EnGen-TTS-M & 280M & 4.35 $\pm$ 0.12 & 768 & 8 & 13 & 0.016 \\
EnGen-TTS-L & 570M & 4.42 $\pm$ 0.08 & 1024 & 16 & 26 & 0.021 \\ \bottomrule
\end{tabular}
\end{adjustbox}
\caption{Ablation Based on Model Size}
\label{table-5}
\end{table*}

For the evaluation of inference performance, all models were tested under identical hardware conditions to ensure consistency and comparability. Specifically, we utilized an NVIDIA T4 GPU equipped with 16 GB of VRAM. Inference was performed with a batch size of 16, and each input had an average token length of 20. Under this setup, the inference speed was approximately 200 milliseconds per batch for all models. This uniform testing environment ensured that the performance metrics reported are directly comparable across all evaluated TTS systems.

\subsection{Ablation}

\begin{table}[h]
\centering
\begin{adjustbox}{max width=\textwidth}
\begin{tabular}{@{}cc@{}}
\toprule
\textbf{Models} & \textbf{MOS} \\ \midrule
Wavgrad \cite{chen2020wavegrad} & 4.36 $\pm$ 0.13 \\
MelGan \cite{kumar2019melgan} & 4.39 $\pm$ 0.11 \\
Univnet \cite{jang2021univnet} & 4.41 $\pm$ 0.12 \\
Hifi-Gans \cite{kong2020hifi} & 4.35 $\pm$ 0.13 \\ \bottomrule
\end{tabular}
\end{adjustbox}
\caption{Ablation With Different Vo-Coder}
\label{tab:my-table}
\end{table}

In our experimentation with different vocoders for generating high-quality audio from latent representations, Univnet emerged as the top performer, achieving a Mean Opinion Score (MOS) of 4.41 $\pm$ 0.12. This slight edge over MelGan (4.39 $\pm$ 0.11) and Wavgrad (4.36 $\pm$ 0.13) suggests its superiority in preserving speech quality and naturalness during waveform reconstruction (see Table \ref{tab:my-table}). While HiFi-GANs (4.35 $\pm$ 0.13) exhibited comparable performance, its slightly lower MOS indicates room for further optimization. Overall, these results highlight the importance of vocoder selection in the Text-to-Speech pipeline, with Univnet demonstrating its potential for creating highly faithful and human-sounding synthetic speech.

Additionally, to explore the impact of model size on both perceptual quality and real-time efficiency, we conducted an ablation study as outlined in Table \ref{table-5}. The table presents results for three variants of our EnGen-TTS model, denoted as EnGen-TTS-S, EnGen-TTS-M, and EnGen-TTS-L, with varying parameters. As model size increases from 87M to 570M, we observe a corresponding improvement in Mean Opinion Score (MOS), indicating enhanced speech quality. Specifically, EnGen-TTS-L achieves a MOS of 4.42 $\pm$ 0.08, outperforming the smaller variants. However, this comes at the cost of increased Real-Time Factor (RTF), with EnGen-TTS-L demonstrating a slightly longer synthesis time (0.021) compared to EnGen-TTS-S (0.014). This trade-off between model size, perceptual quality, and synthesis speed provides valuable insights into tailoring the EnGen-TTS architecture based on specific application requirements and resource constraints.

\begin{table}[t]
\centering 
\begin{adjustbox}{max width=\textwidth}
\begin{tabular}{@{}cc@{}}
\toprule
\textbf{Model} & \textbf{MOS} \\ \midrule
Lgan & 4.30 $\pm$ 0.12 \\
Lmel & 4.32 $\pm$ 0.13 \\
Lgan + Lmel & 4.35 $\pm$ 0.12 \\ \bottomrule
\end{tabular}
\end{adjustbox}
\caption{Ablation based on Loss}
\label{table}
\end{table}

We also conducted an ablation study focusing on the impact of different loss components. Table \ref{table} summarizes the Mean Opinion Scores (MOS) obtained from three model variants: Lgan, Lmel, and the combination of both (Lgan + Lmel). The results indicate that incorporating both the adversarial loss (Lgan) and the mel-spectrogram loss (Lmel) leads to a MOS of 4.35 $\pm$ 0.12, showcasing a marginal improvement over individual losses. This nuanced exploration of loss components provides valuable insights into the synergy between adversarial and mel-spectrogram losses in our training pipeline, contributing to the optimization of our Bahasa TTS model for enhanced speech synthesis quality. Note: For every loss we are always computing $L_{ce}$, without it model can't be trained.

\begin{figure}[t]
    \centering
    \includegraphics[width=0.45\textwidth]{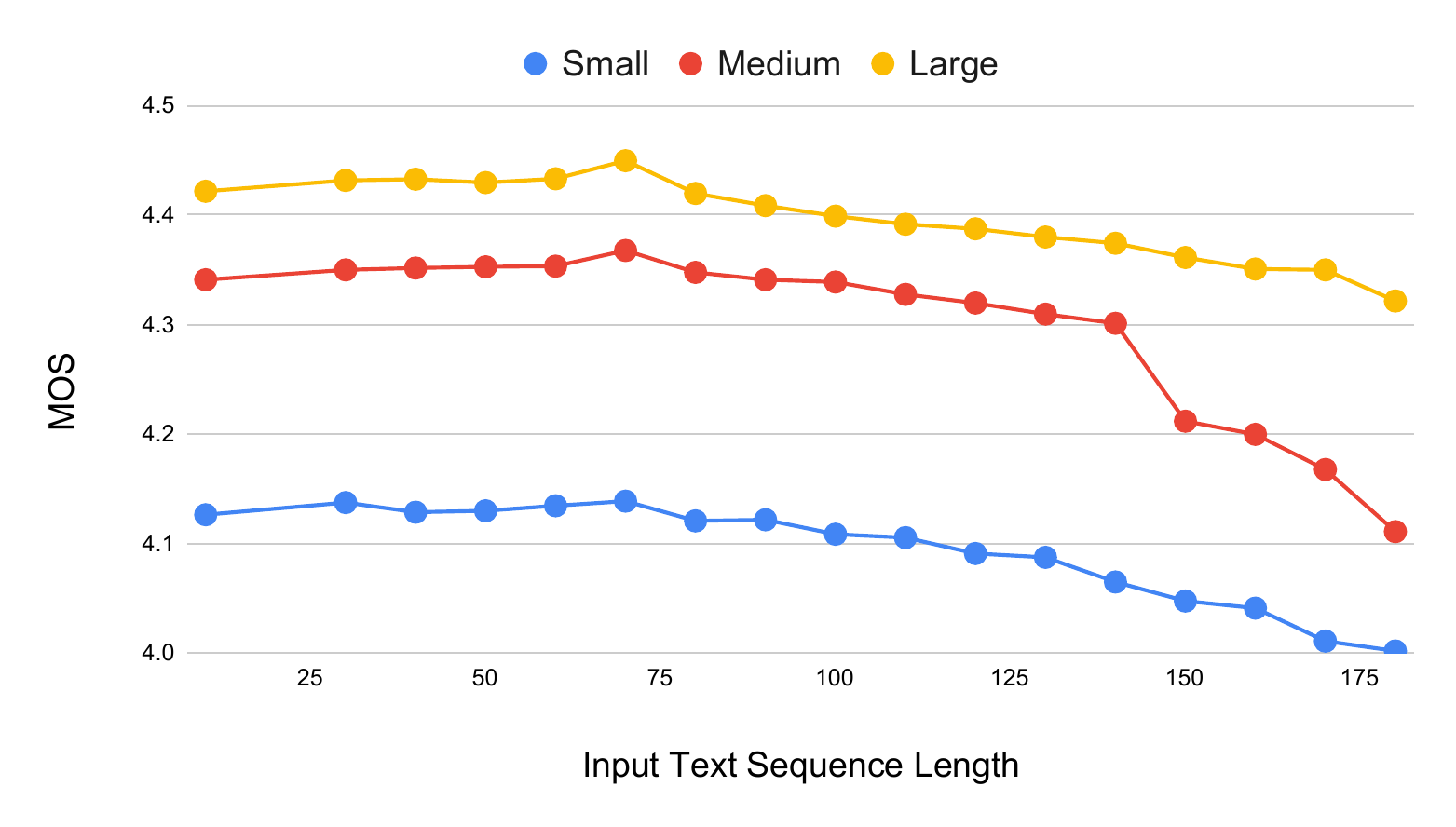}
    \caption{Input text sequence length vs MOS}
    \label{fig:MosvsT}
\end{figure}

We conducted an ablation study on three different model sizes—large, medium, and small—focusing on their performance with varying lengths of input text sequences. Our observations indicate that for text sequences ranging from 5 to 75 tokens, there is minimal variation in the MOS metrics. However, as the sequence length exceeds 75 tokens, we noticed a decline in MOS metrics. This decline correlates with deteriorations in pronunciation and timbre quality of generated speech, along with an increase in the Word Error Rate (WER). As depicted in Figure \ref{fig:MosvsT}, extending the sequence length beyond 100 tokens results in a significant decrease in MOS metrics, likely due to the models' inability to manage longer contexts effectively, leading to catastrophic forgetting which is discuss in the paper \cite{liu2024lost}.


\section{Conclusions}

In conclusion, our study presents a pivotal advancement in Bahasa text-to-speech (TTS) synthesis, combining a richly curated dataset with a groundbreaking model design. Our comprehensive Bahasa TTS dataset, encompassing over 55 hours of audio and 52K recording, is a robust resource, crafted with inputs from proficient voice artists and varied textual content. The introduced model, EnGen-TTS, excels in performance, surpassing conventional benchmarks with its innovative architecture, which includes a multi-task T5 (m-T5) encoder and a neural codec language modeling module, without necessitating extra fine-tuning for Bahasa. This design not only enhances speech synthesis quality but also ensures computation efficiency, establishing a new standard in TTS technology. Our work not only pushes forward the boundaries of Bahasa TTS but also lays the groundwork for future developments in multilingual text-to-speech systems, promising high-quality and diverse linguistic applications.

\section{Limitations}
One limitation of our proposed method is its reliance on audio sampled at 22.05 KHz. This sampling rate is necessitated by the use of Meta's pre-trained Audio Encodec, which requires 22.05 kHz audio data. However, this presents a challenge for applications such as automatic voice calling, where telephony standards typically mandate an 8 kHz sampling rate. The required down-sampling from 22.05 KHz to 8 KHz results in a significant reduction in audio quality, manifesting as "muffled speech" due to the drastic decrease in sampling rate. Future work will focus on enabling high-quality audio generation directly at 8 kHz to better align with telephony requirements without compromising speech clarity.

Another limitation of our method lies in the maximum sequence length used during training, which is capped at 500 audio tokens. This constraint is well-suited for generating high-quality speech for shorter sentences or sentences containing up to 70-80 words. However, when the word count exceeds this limit, the generated speech may exhibit unnatural pauses or occasional missing words. This issue is likely due to catastrophic forgetting of longer contexts. Our future research will focus on increasing the context window up to 2048 audio tokens to better handle larger sentences or paragraphs, thereby improving the naturalness and continuity of generated speech.

\bibliography{custom}

\begin{thebibliography}{29}
\providecommand{\natexlab}[1]{#1}

\bibitem[{Yam(2019)}]{Yamagishi}
 2019.
\newblock English multi-speaker corpus for cstr voice cloning toolkit.
\newblock \emph{https://doi.org/10.7488/ds/2645}.

\bibitem[{Casanova et~al.(2024)Casanova, Davis, G{\"o}lge, G{\"o}knar, Gulea, Hart, Aljafari, Meyer, Morais, Olayemi et~al.}]{casanova2024xtts}
Edresson Casanova, Kelly Davis, Eren G{\"o}lge, G{\"o}rkem G{\"o}knar, Iulian Gulea, Logan Hart, Aya Aljafari, Joshua Meyer, Reuben Morais, Samuel Olayemi, et~al. 2024.
\newblock Xtts: a massively multilingual zero-shot text-to-speech model.
\newblock \emph{arXiv preprint arXiv:2406.04904}.

\bibitem[{Chen et~al.(2020)Chen, Zhang, Zen, Weiss, Norouzi, and Chan}]{chen2020wavegrad}
Nanxin Chen, Yu~Zhang, Heiga Zen, Ron~J Weiss, Mohammad Norouzi, and William Chan. 2020.
\newblock Wavegrad: Estimating gradients for waveform generation.
\newblock \emph{arXiv preprint arXiv:2009.00713}.

\bibitem[{Ito and Johnson(2017)}]{ljspeech17}
Keith Ito and Linda Johnson. 2017.
\newblock The lj speech dataset.
\newblock \url{https://keithito.com/LJ-Speech-Dataset/}.

\bibitem[{Jang et~al.(2021)Jang, Lim, Yoon, Kim, and Kim}]{jang2021univnet}
Won Jang, Dan Lim, Jaesam Yoon, Bongwan Kim, and Juntae Kim. 2021.
\newblock Univnet: A neural vocoder with multi-resolution spectrogram discriminators for high-fidelity waveform generation.
\newblock \emph{arXiv preprint arXiv:2106.07889}.

\bibitem[{Kim et~al.(2020)Kim, Kim, Kong, and Yoon}]{kim2020glow}
Jaehyeon Kim, Sungwon Kim, Jungil Kong, and Sungroh Yoon. 2020.
\newblock Glow-tts: A generative flow for text-to-speech via monotonic alignment search.
\newblock \emph{Advances in Neural Information Processing Systems}, 33:8067--8077.

\bibitem[{Kim et~al.(2021)Kim, Kong, and Son}]{kim2021conditional}
Jaehyeon Kim, Jungil Kong, and Juhee Son. 2021.
\newblock Conditional variational autoencoder with adversarial learning for end-to-end text-to-speech.
\newblock In \emph{International Conference on Machine Learning}, pages 5530--5540. PMLR.

\bibitem[{Kim et~al.(2024)Kim, Lee, Chung, and Cho}]{kim2024clam}
Jaehyeon Kim, Keon Lee, Seungjun Chung, and Jaewoong Cho. 2024.
\newblock Clam-tts: Improving neural codec language model for zero-shot text-to-speech.
\newblock \emph{arXiv preprint arXiv:2404.02781}.

\bibitem[{Kong et~al.(2020{\natexlab{a}})Kong, Kim, and Bae}]{NEURIPS2020_c5d73680}
Jungil Kong, Jaehyeon Kim, and Jaekyoung Bae. 2020{\natexlab{a}}.
\newblock \href {https://proceedings.neurips.cc/paper_files/paper/2020/file/c5d736809766d46260d816d8dbc9eb44-Paper.pdf} {Hifi-gan: Generative adversarial networks for efficient and high fidelity speech synthesis}.
\newblock In \emph{Advances in Neural Information Processing Systems}, volume~33, pages 17022--17033. Curran Associates, Inc.

\bibitem[{Kong et~al.(2020{\natexlab{b}})Kong, Kim, and Bae}]{kong2020hifi}
Jungil Kong, Jaehyeon Kim, and Jaekyoung Bae. 2020{\natexlab{b}}.
\newblock Hifi-gan: Generative adversarial networks for efficient and high fidelity speech synthesis.
\newblock \emph{Advances in Neural Information Processing Systems}, 33:17022--17033.

\bibitem[{Kumar et~al.(2023)Kumar, S~V, Kumar, Khapra, and Nandakumar}]{10096069}
Gokul~Karthik Kumar, Praveen S~V, Pratyush Kumar, Mitesh~M. Khapra, and Karthik Nandakumar. 2023.
\newblock \href {https://doi.org/10.1109/ICASSP49357.2023.10096069} {Towards building text-to-speech systems for the next billion users}.
\newblock In \emph{ICASSP 2023 - 2023 IEEE International Conference on Acoustics, Speech and Signal Processing (ICASSP)}, pages 1--5.

\bibitem[{Kumar et~al.(2019)Kumar, Kumar, De~Boissiere, Gestin, Teoh, Sotelo, De~Brebisson, Bengio, and Courville}]{kumar2019melgan}
Kundan Kumar, Rithesh Kumar, Thibault De~Boissiere, Lucas Gestin, Wei~Zhen Teoh, Jose Sotelo, Alexandre De~Brebisson, Yoshua Bengio, and Aaron~C Courville. 2019.
\newblock Melgan: Generative adversarial networks for conditional waveform synthesis.
\newblock \emph{Advances in neural information processing systems}, 32.

\bibitem[{Kumar et~al.(2024)Kumar, Seetharaman, Luebs, Kumar, and Kumar}]{kumar2024high}
Rithesh Kumar, Prem Seetharaman, Alejandro Luebs, Ishaan Kumar, and Kundan Kumar. 2024.
\newblock High-fidelity audio compression with improved rvqgan.
\newblock \emph{Advances in Neural Information Processing Systems}, 36.

\bibitem[{Liu et~al.(2024)Liu, Lin, Hewitt, Paranjape, Bevilacqua, Petroni, and Liang}]{liu2024lost}
Nelson~F Liu, Kevin Lin, John Hewitt, Ashwin Paranjape, Michele Bevilacqua, Fabio Petroni, and Percy Liang. 2024.
\newblock Lost in the middle: How language models use long contexts.
\newblock \emph{Transactions of the Association for Computational Linguistics}, 12:157--173.

\bibitem[{Lu et~al.(2021)Lu, Wu, Wu, Li, Kang, Liu, and Meng}]{lu2021vaenar}
Hui Lu, Zhiyong Wu, Xixin Wu, Xu~Li, Shiyin Kang, Xunying Liu, and Helen Meng. 2021.
\newblock Vaenar-tts: Variational auto-encoder based non-autoregressive text-to-speech synthesis.

\bibitem[{Oliveira et~al.(2023)Oliveira, Casanova, Junior, Soares, and Galv{\~a}o~Filho}]{oliveira2023cml}
Frederico~S Oliveira, Edresson Casanova, Arnaldo~Candido Junior, Anderson~S Soares, and Arlindo~R Galv{\~a}o~Filho. 2023.
\newblock Cml-tts: A multilingual dataset for speech synthesis in low-resource languages.
\newblock In \emph{International Conference on Text, Speech, and Dialogue}, pages 188--199. Springer.

\bibitem[{Popov et~al.(2021)Popov, Vovk, Gogoryan, Sadekova, and Kudinov}]{popov2021grad}
Vadim Popov, Ivan Vovk, Vladimir Gogoryan, Tasnima Sadekova, and Mikhail Kudinov. 2021.
\newblock Grad-tts: A diffusion probabilistic model for text-to-speech.
\newblock In \emph{International Conference on Machine Learning}, pages 8599--8608. PMLR.

\bibitem[{Schumacher and LaBounty~Jr(2023)}]{schumacher2023enhancing}
Devin Schumacher and Francis LaBounty~Jr. 2023.
\newblock Enhancing suno's bark text-to-speech model: Addressing limitations through meta's encodec and pre-trained hubert.
\newblock \emph{Available at SSRN 4443815}.

\bibitem[{Srivastava et~al.(2020{\natexlab{a}})Srivastava, Mukhopadhyay, K~R, and Jawahar}]{srivastava-etal-2020-indicspeech}
Nimisha Srivastava, Rudrabha Mukhopadhyay, Prajwal K~R, and C~V Jawahar. 2020{\natexlab{a}}.
\newblock \href {https://aclanthology.org/2020.lrec-1.789} {{I}ndic{S}peech: Text-to-speech corpus for {I}ndian languages}.
\newblock In \emph{Proceedings of the Twelfth Language Resources and Evaluation Conference}, pages 6417--6422, Marseille, France. European Language Resources Association.

\bibitem[{Srivastava et~al.(2020{\natexlab{b}})Srivastava, Mukhopadhyay, Prajwal, and Jawahar}]{srivastava2020indicspeech}
Nimisha Srivastava, Rudrabha Mukhopadhyay, KR~Prajwal, and CV~Jawahar. 2020{\natexlab{b}}.
\newblock Indicspeech: text-to-speech corpus for indian languages.
\newblock In \emph{Proceedings of the Twelfth Language Resources and Evaluation Conference}, pages 6417--6422.

\bibitem[{Susladkar et~al.(2023)Susladkar, Gatti, and Yadav}]{susladkar2023slbert}
Onkar Susladkar, Prajwal Gatti, and Santosh~Kumar Yadav. 2023.
\newblock Slbert: A novel pre-training framework for joint speech and language modeling.
\newblock In \emph{ICASSP 2023-2023 IEEE International Conference on Acoustics, Speech and Signal Processing (ICASSP)}, pages 1--5. IEEE.

\bibitem[{Tan et~al.(2024)Tan, Chen, Liu, Cong, Zhang, Liu, Wang, Leng, Yi, He et~al.}]{tan2024naturalspeech}
Xu~Tan, Jiawei Chen, Haohe Liu, Jian Cong, Chen Zhang, Yanqing Liu, Xi~Wang, Yichong Leng, Yuanhao Yi, Lei He, et~al. 2024.
\newblock Naturalspeech: End-to-end text-to-speech synthesis with human-level quality.
\newblock \emph{IEEE Transactions on Pattern Analysis and Machine Intelligence}.

\bibitem[{van~den Oord et~al.(2016)van~den Oord, Dieleman, Zen, Simonyan, Vinyals, Graves, Kalchbrenner, Senior, and Kavukcuoglu}]{45774}
Aäron van~den Oord, Sander Dieleman, Heiga Zen, Karen Simonyan, Oriol Vinyals, Alexander Graves, Nal Kalchbrenner, Andrew Senior, and Koray Kavukcuoglu. 2016.
\newblock \href {https://arxiv.org/abs/1609.03499} {Wavenet: A generative model for raw audio}.
\newblock In \emph{Arxiv}.

\bibitem[{Viswanathan and Viswanathan(2005)}]{VISWANATHAN200555}
Mahesh Viswanathan and Madhubalan Viswanathan. 2005.
\newblock \href {https://doi.org/10.1016/j.csl.2003.12.001} {Measuring speech quality for text-to-speech systems: development and assessment of a modified mean opinion score (mos) scale}.
\newblock \emph{Computer Speech \& Language}, 19(1):55--83.

\bibitem[{Wan et~al.(2018)Wan, Wang, Papir, and Moreno}]{wan2018generalized}
Li~Wan, Quan Wang, Alan Papir, and Ignacio~Lopez Moreno. 2018.
\newblock Generalized end-to-end loss for speaker verification.
\newblock In \emph{2018 IEEE International Conference on Acoustics, Speech and Signal Processing (ICASSP)}, pages 4879--4883. IEEE.

\bibitem[{Wang et~al.(2017)Wang, Skerry-Ryan, Stanton, Wu, Weiss, Jaitly, Yang, Xiao, Chen, Bengio, Le, Agiomyrgiannakis, Clark, and Saurous}]{wang17n_interspeech}
Yuxuan Wang, R.J. Skerry-Ryan, Daisy Stanton, Yonghui Wu, Ron~J. Weiss, Navdeep Jaitly, Zongheng Yang, Ying Xiao, Zhifeng Chen, Samy Bengio, Quoc Le, Yannis Agiomyrgiannakis, Rob Clark, and Rif~A. Saurous. 2017.
\newblock \href {https://doi.org/10.21437/Interspeech.2017-1452} {{Tacotron: Towards End-to-End Speech Synthesis}}.
\newblock In \emph{Proc. Interspeech 2017}, pages 4006--4010.

\bibitem[{Xue et~al.(2021)Xue, Constant, Roberts, Kale, Al-Rfou, Siddhant, Barua, and Raffel}]{xue-etal-2021-mt5}
Linting Xue, Noah Constant, Adam Roberts, Mihir Kale, Rami Al-Rfou, Aditya Siddhant, Aditya Barua, and Colin Raffel. 2021.
\newblock \href {https://doi.org/10.18653/v1/2021.naacl-main.41} {m{T}5: A massively multilingual pre-trained text-to-text transformer}.
\newblock In \emph{Proceedings of the 2021 Conference of the North American Chapter of the Association for Computational Linguistics: Human Language Technologies}, pages 483--498, Online. Association for Computational Linguistics.

\bibitem[{Zhang et~al.(2023)Zhang, Zhou, Wang, Chen, Wu, Liu, Chen, Liu, Wang, Li et~al.}]{zhang2023speak}
Ziqiang Zhang, Long Zhou, Chengyi Wang, Sanyuan Chen, Yu~Wu, Shujie Liu, Zhuo Chen, Yanqing Liu, Huaming Wang, Jinyu Li, et~al. 2023.
\newblock Speak foreign languages with your own voice: Cross-lingual neural codec language modeling.
\newblock \emph{arXiv preprint arXiv:2303.03926}.

\bibitem[{Łańcucki(2021)}]{9413889}
Adrian Łańcucki. 2021.
\newblock \href {https://doi.org/10.1109/ICASSP39728.2021.9413889} {Fastpitch: Parallel text-to-speech with pitch prediction}.
\newblock In \emph{ICASSP 2021 - 2021 IEEE International Conference on Acoustics, Speech and Signal Processing (ICASSP)}, pages 6588--6592.

\end{thebibliography}


\end{document}